\documentclass[letterpaper, 10 pt, conference]{ieeeconf}
\usepackage[utf8]{inputenc}
\usepackage{graphicx}
\usepackage{ulem,tikz,pgfplots}
\usepackage{grffile}
\usepackage{amsmath}
\usepackage{gensymb}
\usepackage{url}
\usepackage{array}
\usepackage{verbatim}
\usepackage{booktabs}
\usepackage[flushleft]{threeparttable}
\usepackage[shortcuts]{extdash}
\usepackage{fancyref}

\pgfplotsset{compat=newest}
\usetikzlibrary{plotmarks}
\usetikzlibrary{arrows.meta}
\usepgfplotslibrary{patchplots}
\pgfplotsset{plot coordinates/math parser=false}
\newlength\figureheight
\newlength\figurewidth

\usetikzlibrary{shapes,arrows,calc,intersections,positioning,fit}

\newlength{\dotSymRad}
\newlength{\hammerlength}
\newlength{\estoplength}
\newcommand{\angleThetaH}{140} 
\newcommand{\angleThetaE}{150} 

\newcommand{\drawPoint}[4]{
	\coordinate (point#1) at (#2);
	\draw [fill=black,radius=\dotSymRad] (point#1) circle;
	\node[label] (node#1)[node distance=.5em,#4=of point#1]{#3};
}

\newcommand{\tikzAngleOfLine}{\tikz@AngleOfLine}
\def\tikz@AngleOfLine(#1)(#2)#3{%
	\pgfmathanglebetweenpoints{%
		\pgfpointanchor{#1}{center}}{%
		\pgfpointanchor{#2}{center}}
	\pgfmathsetmacro{#3}{\pgfmathresult}%
}

\newcommand{\drawAngle}[5]{
	\tikzAngleOfLine(#1)(#2){\angleLineOA}
	\tikzAngleOfLine(#1)(#3){\angleLineOB}
	\draw[->] ($(#1)!#4!(#2)$) arc (\angleLineOA:\angleLineOB:#4);
	\coordinate (n1) at ($(#1)!#4!(#2)$);
	\coordinate (n2) at ($(#1)!#4!(#3)$);
	\coordinate (n3) at ($(n1)!0.5!(n2)$);
	\coordinate (n4) at ($(#1)!3em+#4!(n3)$);
	\node at (n4) {#5};
}

\IEEEoverridecommandlockouts 

\newcommand{\plotheightA}{.135\textheight}

\begin{document}
	\title{\LARGE \bf
		Towards Teleoperation with Human-like Dynamics: Human Use of Elastic Tools}
	\author{Manuel Aiple$^{1}$,~\IEEEmembership{Student Member,~IEEE,} and André Schiele$^{1}$%
		\thanks{$^{1}$M. Aiple and A. Schiele are affiliated with the Department of BioMechanical Engineering, Delft University of Technology, Delft, The Netherlands.\protect\\%
			{\tt\small \{m.aiple,a.schiele\}@tudelft.nl.}%
		}
	}
	
	\maketitle
	\thispagestyle{empty}
	\pagestyle{empty}
	
	\begin{abstract}
		Variable stiffness actuators undergo lower peak force in contacts compared to their rigid counterparts, and are thus safer for human-robot interaction.
		Furthermore, they can store	energy in their elastic element and can release it later to achieve human-like dynamic movements.
		However, it is not clear how to integrate them in teleoperator systems so that they can be controlled intuitively by a human.
		We performed an experiment to study human use of elastic tools to determine how a teleoperator system with an elastic slave would	need to be designed.
		For this, we had 13 untrained participants hammer with an elastic tool under different stiffness conditions, asking them to try to	find the best timing for a backward-forward swing motion in order to achieve the strongest impact.
		We found that the participants generally executed the task efficiently after a few
		trials and they converged to very similar solutions.
		The stiffness influenced the performance slightly, a stiffness between 2.3~Nm/rad and 4.1~Nm/rad showing the best results.
		We conclude that humans intuitively know how to efficiently use elastic tools for hammering type tasks.
		This could facilitate the control of teleoperator systems with elastic slave manipulators for tasks requiring explosive movements like hammering.
	\end{abstract}
	
	\section{Introduction}
	Teleoperation promises to combine the advantages of a human operator in terms of flexibility and planning with the benefits of a robot in terms of robustness to harsh environments (e.g., radioactively contaminated areas, subsea, space). Extensive research has been done to find methods and technologies to efficiently couple the motion of the operator on the master side with the motion of the robot on the slave side, addressing issues like stability \cite{lawrence_stability_1993}, transparency, time-delay \cite{pinto_rebelo_robust_2015}, sharing of control \cite{groten_shared_2010}. However, with the robots and input devices of today, the operator is mostly limited to slow motions with low dynamics.
	
	Series Elastic Actuators (SEAs) and variable stiffness actuators (VSAs) have increasingly become a topic of interest over the last years. Vanderborght et al. give a good overview of the existing systems \cite{van_ham_compliant_2009}. Unlike active compliance, which emulates a spring effect by force measurement and control, SEAs and VSAs are equipped with a physical spring on the output of the actuator. For SEAs with constant stiffness, this is a mechanical spring, whereas VSAs incorporate a mechanism to change their effective stiffness. This gives SEAs and VSAs true mechanical compliance and higher admittance at high frequencies than actuators with active compliance, resulting in reduced forces in highly dynamic situations. In particular, they are more robust to impacts and safer for humans to interact with.
	
	Furthermore, SEAs and VSAs can store energy in the spring element to release it again later. Previous studies focused on how to exploit this to make autonomous robots more energy efficient in high dynamic tasks like walking and throwing \cite{braun_optimal_2012}\cite{garabini_optimality_2011}.
	Our research aims at developing methods and technologies for teleoperation of dynamic real-life human tasks, like hammering, jolting, throwing, etc., using VSAs.
	But, for teleoperation with elastic tools, there are competing objectives between following the human input accurately to achieve maximum transparency and achieving the best task performance by following an optimal control strategy. In order to resolve this conflict, knowledge about human task performance using flexible tools is required.
	If humans have an intuitive understanding of how to optimally use elastic tools, teleoperator systems should be optimized for maximum transparency also for teleoperation with flexible tools. On the other hand, if human performance is bad with elastic systems, a shared control approach based on optimal control of the slave might be better.
	
	This paper examines how humans interact with elastic tools for maximum output velocity tasks. Maximum output velocity tasks are tasks like throwing and hammering, aimed at giving the most kinetic energy to an object with as little input energy as possible. We picked this class of tasks for this experiment as it makes use of SEAs' advantage compared to rigid actuators, their capacity to store energy in their spring when switching from a backward motion to forward motion and then releasing it as kinetic energy. Optimal control strategies have been proposed for SEAs and VSAs executing maximum output velocity tasks \cite{haddadin_optimal_2011}, and output velocities of 272\% of the maximum motor speed have been measured with VSAs \cite{wolf_new_2008}.
	
	We took one step back from the teleoperation scenario to exclude effects coming from the teleoperator system (e.g., time-delay, missing fidelity in force-feedback rendering, dynamic constraints of the slave) and performed an experiment with direct use of an elastic tool, asking participants to hit a target as hard as possible with an elastic hammer (section \ref{sec:method}). We analyzed the task performance as a function of the stiffness of the elastic hammer, the number of trials and the motion profile (section \ref{sec:analysis}). Our results suggest that humans can intuitively use an elastic hammer exploiting the spring effect (section \ref{sec:results}). We conclude that this might be beneficial for future teleoperation systems with VSAs (section \ref{sec:conclusion}) .
	
	\section{Method}
	\label{sec:method}
\begin{figure}
		\input{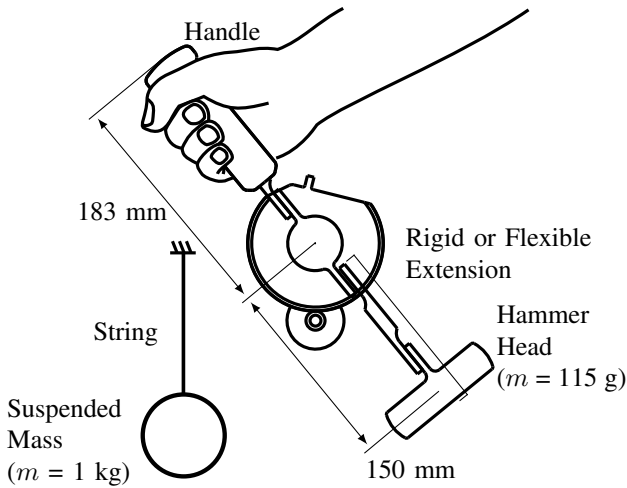}
		\caption{Experimental setup. The motion is constrained to rotation around one axis. The extension can be exchanged to allow for different stiffness of the hammer. The movement of the suspended mass shows to the participants how hard they hit the target.}
		\label{fig:setup}
	\end{figure}

	For this experiment, we modified the one degree of freedom teleoperator system designed by Rebelo et al. \cite{pinto_rebelo_robust_2015} for direct interaction with an elastic tool. Fig.~\ref{fig:setup} shows a diagram of the setup.
	We used only one unit and decoupled the motor from the output axis by removing the wire from the capstan gear. We replaced the original end-effector by a handle and a hammer.
	The hammer had two parts: the hammer head and an interchangeable extension. The latter could be completely rigid, in which case the tool was a rigid hammer, or a flexible extension realized with a leaf-spring, in which case the tool was an elastic hammer.
	
	We mounted the hammer hanging down, so that the spring was stable in neutral position with the handle in  vertical position. Participants could execute hammering motions by moving the handle with their right hand. The setup constrained the motion to one degree of freedom.
	Therefore, the motion was quite different from the hammering participants might have been accustomed to. In addition, the backswing was a pushing motion for the participants and the forward swing a pulling motion. Although one might consider this as making the experimental setup unrealistic, we thought that this was beneficial as it reduced possible effects from previously acquired skills of participants.
	We assumed that participants had a general understanding of elastic tools, but no knowledge about the setup in particular.
	
	We asked 13 participants (8 male, 5 female, 12 right-handed, 1 ambidextrous, age 21-41) to perform a hammering task on the setup. The participants were told that the goal of this experiment was to study how intuitive it is for humans to interact with an elastic tool. The participants were instructed to hit the target with the hammer by executing one backward swing followed by a forward swing. Before every hit, the hammer should be in the resting position with the hammer head slightly touching the target in order to ensure the hammer head was at rest before the motion. The hammer should hit the target only once per movement and not repeatedly. Comfortable and similar effort should be used for all hits, focusing on the timing for switching from backward swing to forward swing (related to the input frequency explained below) to achieve the maximum impact. The experiment was approved by the human research ethics committee of TU Delft and all participants gave written informed consent before participation.
	
	The mass of the hammer head and its distance to the center of rotation were constant at 115~g and 150~mm. We used five different extensions: the rigid extension and four leaf-springs of different thickness and stiffness: 0.4~mm (0.62~Nm/rad), 0.6~mm (2.3~Nm/rad), 0.8~mm (4.1~Nm/rad), and 1.0~mm (11~Nm/rad). In the following we will use ``Rigid'' or the spring thickness as names for the experiment conditions. 
	As a first approximation, the combination of leaf-spring and hammer head can be modeled as a mass-spring system. Its resonance frequency $f_0$ depends only on the rotational spring constant $\kappa$  and the moment of inertia of the hammer head $I$ according to the following equation: $f_0 = \frac{1}{2 \pi} \sqrt{\frac{\kappa}{I}}$.
	If the input excitation through the handle is close to the resonance frequency, mechanical resonance occurs. In this case, the hammer head velocity amplitude is greater than the handle velocity amplitude.  Thus, for each condition, the participants needed to adapt to the resonance frequency of the elastic tool to achieve a high peak output velocity with a low peak input velocity corresponding to a high impact with little effort. However, besides the instructions explained above, the participants were not taught a specific input profile, but had to adjust in the way they deemed the best for moving the suspended mass compared to the effort they put into the strike.
	
	For every condition, the participants had a learning phase of approximately 100~trials and a performance phase of approximately 20~trials. During the learning phase, the participants were asked to try to improve their performance by experimenting with different timings (i.e., executing the motion at different speeds). During the performance phase, the participants were asked to use the timing which they had found to be the best during the learning phase.
	We chose 100 trials, as experience from the pilot experiments showed that this was a trade-off allowing enough time to get used to a condition without getting tired. After every phase there was a short interruption to download the measurement data, and between every condition there was a short break to change the hammer extension.
	We chose a different sequence of conditions per participant in order to compensate for learning effects from one condition to the next.
	
\begin{figure}[!t]
		\centering
		\includegraphics[height=.23\textheight]{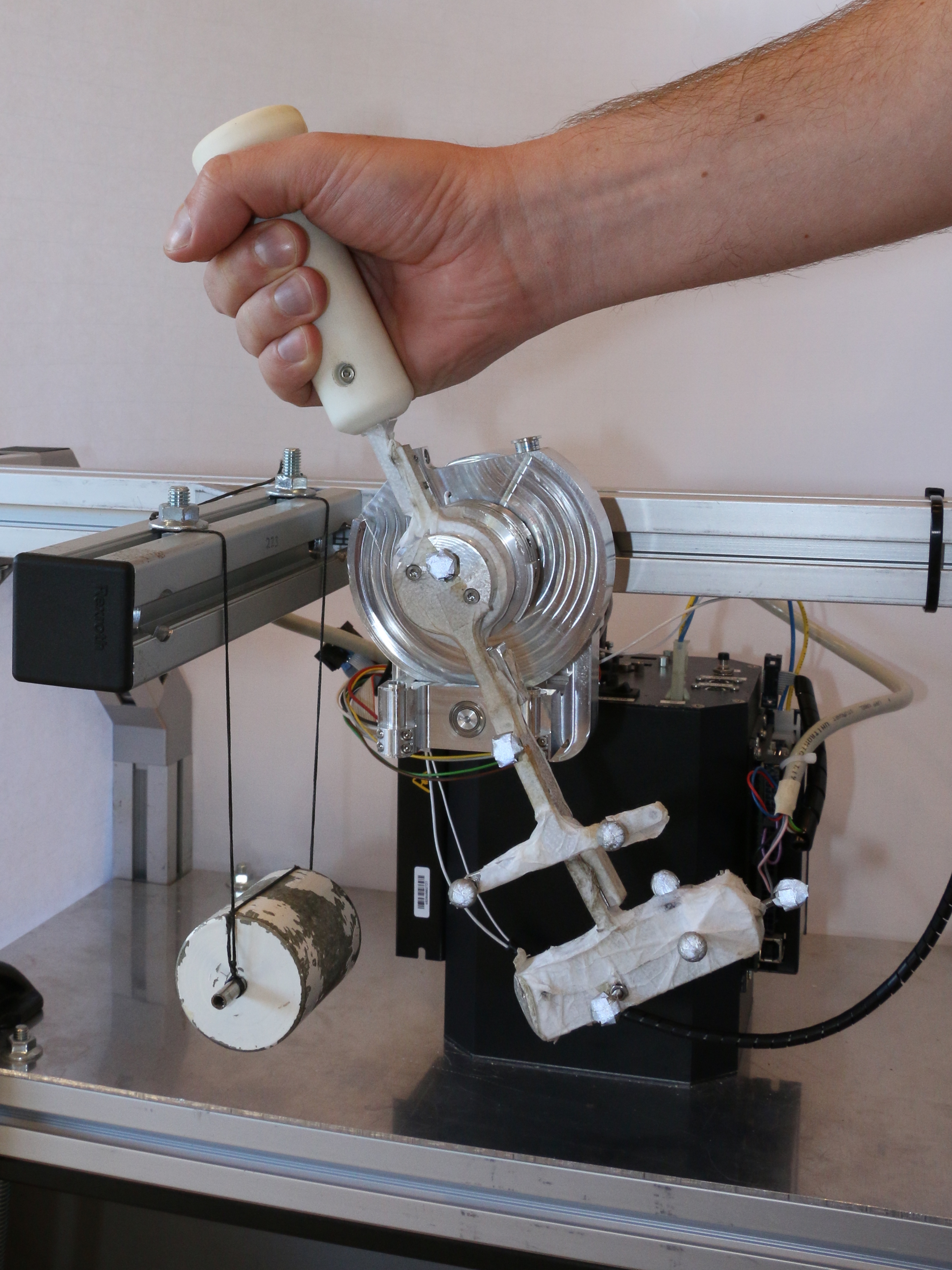}
		\caption{Participant holding the handle of the setup. The reflective markers on the handle and on the hammer head allow position tracking with the motion capture system. }
		\label{fig:photo-setup}
	\end{figure}

	We measured the handle position and the hammer position using a Vicon motion capture system running at a frame rate of 1~kHz. For this we attached passive reflective markers on an extension to the handle and at the hammer head (cf. Fig.~\ref{fig:photo-setup}).%
	
	\section{Analysis}
	\label{sec:analysis}
\begin{figure}[!t]
		\centering
		\begin{tikzpicture}[scale=0.2]
\drawPoint{C}{0,0}{C}{above}
\drawPoint{R}{$(pointC)+(15cm,-2cm)$}{R}{right}
\drawPoint{H}{$(pointC)+({\hammerlength*sin(\angleThetaH)},{\hammerlength*cos(\angleThetaH)})$}{H}{below}
\drawPoint{E}{$(pointC)+({\estoplength*sin(\angleThetaE)},{\estoplength*cos(\angleThetaE)})$}{E}{below}
\coordinate (pointDown) at ($(pointC)-(0,\hammerlength)$);
\coordinate (pointRx) at ($(pointR)-(3cm,0)$);
\coordinate (pointRy) at ($(pointR)+(1.5cm,-2cm)$);
\coordinate (pointRz) at ($(pointR)+(0,3cm)$);
\draw[line width=0.2pt] (pointC) -- (pointDown);
\draw[name path=lineH] (pointC) -- (pointH);
\draw[name path=lineE] (pointC) -- (pointE);
\drawAngle{pointC}{pointDown}{pointE}{5cm}{$\theta_e$}
\drawAngle{pointC}{pointDown}{pointH}{8cm}{$\theta_h$}
\draw[>=latex,->] (pointR) -- (pointRx);
\draw[>=latex,->] (pointR) -- (pointRy);
\draw[>=latex,->] (pointR) -- (pointRz);
\node at (pointRx)[below left] {$x$};
\node at (pointRy)[below right] {$y$};
\node at (pointRz)[above right] {$z$};

\end{tikzpicture}
		\caption{Mapping from motion capture system coordinates $x$,$y$,$z$ in the reference frame $R$ to rotation angles of the handle $\theta_e$ and the hammer head $\theta_h$ around the rotation axis center $C$. The points $E$ and $H$ are the positions of the objects defined by the markers on the handle extension and the hammer head, respectively.}
		\label{fig:geometry}
	\end{figure}
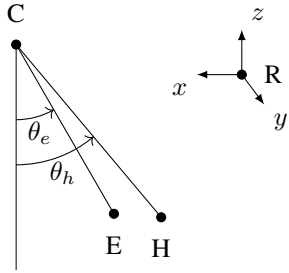

\begin{figure}[!t]
		\centering
		\setlength{\figurewidth}{.8\columnwidth}
		\setlength{\figureheight}{.15\textheight}
		\input{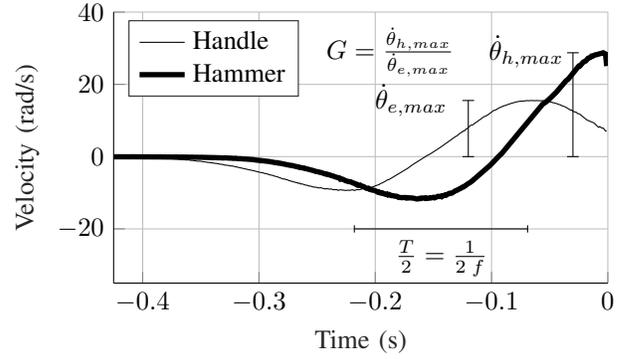}
		\caption{Typical velocity profile for one hammer strike with the metrics used for analysis, showing the handle velocity $\dot{\theta}_{e}$ (thin line), the hammer velocity $\dot{\theta}_{h}$ (thick line), the peak velocities $\dot{\theta}_{e,max}$ and $\dot{\theta}_{h,max}$, and the half-period of the input velocity $\frac{T}{2}$ from minimum to maximum of the handle velocity.}
		\label{fig:measures-explanations}
	\end{figure}

	The markers fixed to the handle and the hammer head defined two positions in the motion capture system: position $E$, coupled to the handle, and position $H$, coupled to the hammer head. We converted the Cartesian coordinates $x$, $y$, $z$ measured by the motion capture system into polar coordinates $\theta_e$ and $\theta_h$ by mapping them to two circles around a common center $C$ and lying in the same plane (cf. Fig.~\ref{fig:geometry}).%
	This produced two one-dimensional signals for analysis. We segmented the time sequences to isolate the individual hammer strikes and synchronized the segments to have the impact with the suspended mass at time 0~s. For the following calculations, we cropped the segments to the time -425~ms up to 0~s. For every segment, we calculated the gain $G$ defined by the peak hammer head velocity $\dot{\theta}_{h,max}$ divided by the peak handle velocity $\dot{\theta}_{e,max}$, and estimated the input frequency $f$ from the half period between minimum and maximum handle velocity (cf. Fig.~\ref{fig:measures-explanations}).%
	
	As explained above, we modeled the elastic hammer as a mass-spring system with the handle velocity as input and the hammer head velocity as output. We extracted the second-order transfer function for the different extensions through system identification to determine the resonance frequency $f_0$ and damping ratio $\zeta$ (table~\ref{tbl:system-identification}). For the system identification we used a separate measurement with manual input at different frequencies with the hammer head able to move freely (suspended mass removed).%
	\begin{table}
		\centering
		\begin{threeparttable}
			\caption{System identification results for the transfer function $H(s) = \frac{\dot{\theta}_h}{\dot{\theta}_e} = \frac{1}{1 + \frac{2 \zeta}{\omega_0} s + \frac{1}{\omega_0^2} s^2}$}
			\label{tbl:system-identification}
			
			\begin{tabular}{lrrr}
				\toprule
				Condition & $ f_0 = \frac{\omega_0}{2 \pi}$ & $\zeta$ \\
				\midrule
				0.4 mm & 2.98 Hz & $23 \cdot 10^{-3}$ \\
				0.6 mm & 4.82 Hz & $32 \cdot 10^{-3}$ \\
				0.8 mm & 6.93 Hz & $40 \cdot 10^{-3}$ \\
				1.0 mm & 9.89 Hz & $17 \cdot 10^{-3}$ \\
				\bottomrule
			\end{tabular}
		\end{threeparttable}
	\end{table}
	Based on this transfer function, we calculated the reference gain $G_{ref}$ with a theoretical reference velocity profile input, which was a bang-bang type profile with half a period of back-swing followed by half a period of forward swing according to Haddadin et al. \cite{haddadin_optimal_2011}.%
	\begin{figure}[!t]
		\centering
		\setlength{\figurewidth}{.8\columnwidth}
		\setlength{\figureheight}{\plotheightA}
		\input{velocity-profile-median-performance.tikz}
		\caption{Median normalized velocity profiles of input (handle) and output (hammer head) for the performance phase trials of the different conditions (1 = peak input velocity of the respective trial). Thin lines represent the respective input velocity and thick lines the output velocity. For the rigid condition, input and output velocity are identical. The input velocity profile is similar for the different conditions but executed slower for softer springs and faster for stiffer springs.}
		\label{fig:velocity-profile-median-performance}
		\vspace{1em}
		\setlength{\figurewidth}{.8\columnwidth}
		\setlength{\figureheight}{\plotheightA}
		\input{human-input-simulated.tikz}
		\caption{Simulated system response to a human input profile. Thin lines are used for the input and thick lines for the output velocity (1 = peak input velocity of the respective trial).}
		\label{fig:human-input-simulated}
		\vspace{1em}
		\setlength{\figurewidth}{.8\columnwidth}
		\setlength{\figureheight}{\plotheightA}
%
%
\definecolor{mycolor1}{rgb}{0.00000,0.44700,0.74100}%
\definecolor{mycolor2}{rgb}{0.85000,0.32500,0.09800}%
\definecolor{mycolor3}{rgb}{0.92900,0.69400,0.12500}%
\definecolor{mycolor4}{rgb}{0.49400,0.18400,0.55600}%
\begin{tikzpicture}

\begin{axis}[%
width=0.951\figurewidth,
height=\figureheight,
at={(0\figurewidth,0\figureheight)},
scale only axis,
xmin=-0.425,
xmax=0,
xlabel style={font=\color{white!15!black}},
xlabel={Time (s)},
ymin=-2,
ymax=3,
ylabel style={font=\color{white!15!black}},
ylabel={Normalized velocity (-)},
axis background/.style={fill=white},
axis x line*=bottom,
axis y line*=left,
xmajorgrids,
ymajorgrids,
legend style={at={(0.03,0.97)}, anchor=north west, legend cell align=left, align=left, draw=white!15!black}
]
\addplot [color=mycolor1]
  table[row sep=crcr]{%
-0.425	0\\
-0.336	0\\
-0.335	-0.554383782335378\\
-0.168	-0.554383782335378\\
-0.167	0.554383782335378\\
-0.00100000000000033	0.554383782335378\\
};
\addlegendentry{0.4 mm}

\addplot [color=mycolor1, line width=2.0pt, forget plot]
  table[row sep=crcr]{%
-0.425	0\\
-0.333	-0.000387825426495514\\
-0.33	-0.00242029805914479\\
-0.327	-0.00618348932078261\\
-0.323	-0.0138642006046612\\
-0.319	-0.0245383723450319\\
-0.314	-0.0419822296848307\\
-0.308	-0.0686974710775465\\
-0.301	-0.107364369653606\\
-0.293	-0.160563564464773\\
-0.284	-0.230386321635337\\
-0.273	-0.327149118836198\\
-0.257	-0.482057894086544\\
-0.229	-0.754794777704655\\
-0.217	-0.857200227642813\\
-0.207	-0.930355669366268\\
-0.199	-0.979114072399595\\
-0.192	-1.01376391118847\\
-0.186	-1.03706870686149\\
-0.181	-1.05177537955621\\
-0.176	-1.06207777931031\\
-0.172	-1.06709987340819\\
-0.169	-1.06897506736953\\
-0.167	-1.06932316309025\\
-0.165	-1.0681747443558\\
-0.163	-1.06475884906339\\
-0.16	-1.05540351006245\\
-0.157	-1.04100960005708\\
-0.153	-1.0140833612181\\
-0.148	-0.968259112357318\\
-0.143	-0.909350576357117\\
-0.137	-0.822201116336033\\
-0.13	-0.699357951343548\\
-0.122	-0.53382241656145\\
-0.112	-0.294938944891582\\
-0.0990000000000002	0.055203167866773\\
-0.0510000000000002	1.39448891866683\\
-0.0410000000000004	1.61592276735273\\
-0.0330000000000004	1.76542720888502\\
-0.0260000000000002	1.87334150758569\\
-0.02	1.94744246497228\\
-0.0150000000000001	1.99556454372684\\
-0.0100000000000002	2.03090118414608\\
-0.00600000000000023	2.04979161363288\\
-0.00300000000000011	2.05843449507166\\
-0.00100000000000033	2.06155593334169\\
};
\addplot [color=mycolor2]
  table[row sep=crcr]{%
-0.425	0\\
-0.208	0\\
-0.207	-0.646463356642973\\
-0.104	-0.646463356642973\\
-0.103	0.646463356642973\\
-0.00100000000000033	0.646463356642973\\
};
\addlegendentry{0.6 mm}

\addplot [color=mycolor2, line width=2.0pt, forget plot]
  table[row sep=crcr]{%
-0.425	0\\
-0.206	-0.000296904061765524\\
-0.204	-0.00266699900863543\\
-0.202	-0.00738956694672011\\
-0.199	-0.0188238722008496\\
-0.195	-0.0419850489758433\\
-0.191	-0.0738091424470428\\
-0.186	-0.124947494935328\\
-0.18	-0.201232769515453\\
-0.172	-0.323745887774386\\
-0.161	-0.517992568221855\\
-0.139	-0.913555089063978\\
-0.131	-1.03298052106016\\
-0.124	-1.11772244263131\\
-0.118	-1.17282549991488\\
-0.113	-1.20510994390468\\
-0.109	-1.2215486965159\\
-0.106	-1.22827588713013\\
-0.104	-1.23007433925679\\
-0.103	-1.23016761683402\\
-0.102	-1.22913091786972\\
-0.1	-1.22190277486277\\
-0.0980000000000003	-1.20784887790984\\
-0.0950000000000002	-1.17414790931246\\
-0.0910000000000002	-1.10625839745542\\
-0.0870000000000002	-1.0132559748314\\
-0.0820000000000003	-0.864093225327241\\
-0.0760000000000001	-0.64190026926116\\
-0.0680000000000001	-0.285490277919572\\
-0.0570000000000004	0.278956553150821\\
-0.0350000000000001	1.42657546098407\\
-0.0270000000000001	1.77247699082834\\
-0.02	2.01761080453133\\
-0.0140000000000002	2.17672602621336\\
-0.00900000000000034	2.26969066571205\\
-0.00500000000000034	2.31678689629844\\
-0.00200000000000022	2.33584136989083\\
-0.00100000000000033	2.33907372679672\\
};
\addplot [color=mycolor3]
  table[row sep=crcr]{%
-0.425	0\\
-0.145	0\\
-0.144	-0.690057561230386\\
-0.0730000000000004	-0.690057561230386\\
-0.0720000000000001	0.690057561230386\\
-0.00100000000000033	0.690057561230386\\
};
\addlegendentry{0.8 mm}

\addplot [color=mycolor3, line width=2.0pt, forget plot]
  table[row sep=crcr]{%
-0.425	0\\
-0.143	-0.000653153128112649\\
-0.141	-0.00585728417718023\\
-0.139	-0.0161914442208668\\
-0.136	-0.0410522933472786\\
-0.132	-0.0907856898102231\\
-0.127	-0.1770418196139\\
-0.121	-0.309790860151399\\
-0.112	-0.548233443397395\\
-0.0950000000000002	-1.00533064148769\\
-0.089	-1.13188415322686\\
-0.0840000000000001	-1.21294532752026\\
-0.0800000000000001	-1.25928685909965\\
-0.0770000000000004	-1.28244087037712\\
-0.0740000000000003	-1.29531406071309\\
-0.0720000000000001	-1.2981226331373\\
-0.0710000000000002	-1.29648941161303\\
-0.0690000000000004	-1.28198815101569\\
-0.0670000000000002	-1.25269731470803\\
-0.0640000000000001	-1.18176957409965\\
-0.0600000000000001	-1.03934910220023\\
-0.0550000000000002	-0.791779843575811\\
-0.0490000000000004	-0.410189975910693\\
-0.04	0.27610792288686\\
-0.0230000000000001	1.59396509304735\\
-0.0170000000000003	1.95949821156761\\
-0.012	2.19397857609075\\
-0.00800000000000001	2.32831848443561\\
-0.00500000000000034	2.39567482464072\\
-0.00200000000000022	2.43342293876266\\
-0.00100000000000033	2.43935419584328\\
};
\addplot [color=mycolor4]
  table[row sep=crcr]{%
-0.425	0\\
-0.102	0\\
-0.101	-0.711547498842229\\
-0.0510000000000002	-0.711547498842229\\
-0.0500000000000003	0.711547498842229\\
-0.00100000000000033	0.711547498842229\\
};
\addlegendentry{1.0 mm}

\addplot [color=mycolor4, line width=2.0pt, forget plot]
  table[row sep=crcr]{%
-0.425	0\\
-0.101	0\\
-0.1	-0.00137118066089537\\
-0.0980000000000003	-0.0122906550479023\\
-0.0960000000000001	-0.0339151082568843\\
-0.093	-0.0855487539419642\\
-0.089	-0.187020565053415\\
-0.0840000000000001	-0.356742079500666\\
-0.0760000000000001	-0.686537941796153\\
-0.0660000000000003	-1.08942617043392\\
-0.0610000000000004	-1.24274792855572\\
-0.0570000000000004	-1.32830528370058\\
-0.0540000000000003	-1.36740949195603\\
-0.052	-1.38080232185658\\
-0.0510000000000002	-1.38362060681134\\
-0.0500000000000003	-1.38384335171319\\
-0.0490000000000004	-1.37873311967031\\
-0.0470000000000002	-1.34445604950842\\
-0.0450000000000004	-1.27871311913092\\
-0.0420000000000003	-1.12374499101599\\
-0.0380000000000003	-0.821524286125326\\
-0.0330000000000004	-0.318525813456948\\
-0.024	0.781179159931738\\
-0.0150000000000001	1.8380407757366\\
-0.0100000000000002	2.28595442854445\\
-0.00600000000000023	2.53434770684303\\
-0.00300000000000011	2.64657870788648\\
-0.00100000000000033	2.68400360206447\\
};
\end{axis}
\end{tikzpicture}%
		\caption{Simulated system response to a square input profile of an ideal velocity source for reference gain calculation. Thin lines are used for the input and thick lines for the output velocity.}
		\label{fig:square-input-simulated}
	\end{figure}%
	
	Figs.~\ref{fig:velocity-profile-median-performance} to \ref{fig:square-input-simulated} show the qualitative difference between the velocity profiles and system responses.
	Fig. \ref{fig:velocity-profile-median-performance} shows the measured input (handle) and output (hammer head) velocity profiles after normalization and taking the median of all participant performance phase trials. The median velocity profile is suitable as a representation of a typical velocity profile as the participants adopted very similar profiles at the end of the learning phase.
	Fig.~\ref{fig:human-input-simulated} visualizes the simulated system response to the human input velocity profile, showing a good match between simulation and measurement data compared to Fig.~\ref{fig:velocity-profile-median-performance}.
	Fig.~\ref{fig:square-input-simulated} shows the results of the simulated system response to the reference input velocity profiles. We can see that the difference in timing is more prominent for the reference profiles than for the human input profiles. The reference profiles have much shorter back swing and forward swing phases for stiffer springs than for softer springs whereas the human input profiles are fairly similar (this is detailed quantitatively in the results section).
	
	We used the following metrics to evaluate the performance of the participants:
	\begin{enumerate}
		\item The peak hammer velocity $\dot{\theta}_{h,max}$ shows which condition gives the best results overall.
		\item The gain $G$ shows which condition is the most efficient in terms of output per input.
		\item The relative gain $\eta = \frac{G}{G_{ref}}$ indicates how well the participants performed compared to the bang-bang type input profile.
		\item The estimated input frequency $f$ shows how the participants adapted to the stiffness of the hammer for the different conditions (values closer to the resonance frequency meaning better adaptation).
		\item The relative frequency error $e = \frac{f-f_0}{f_0}$ indicates how close to the resonance frequency the participants excited the mass-spring system for the different conditions.
	\end{enumerate}

	\section{Results}
	\label{sec:results}
\begin{figure}[!t]
		\centering
		\setlength{\figurewidth}{.9\columnwidth}
		\setlength{\figureheight}{\plotheightA}
%
%
\definecolor{mycolor1}{rgb}{0.00000,0.44700,0.74100}%
\definecolor{mycolor2}{rgb}{0.85000,0.32500,0.09800}%
\definecolor{mycolor3}{rgb}{0.92900,0.69400,0.12500}%
\definecolor{mycolor4}{rgb}{0.49400,0.18400,0.55600}%
\definecolor{mycolor5}{rgb}{0.46600,0.67400,0.18800}%
\definecolor{mycolor6}{rgb}{0.30100,0.74500,0.93300}%
\definecolor{mycolor7}{rgb}{0.63500,0.07800,0.18400}%
\begin{tikzpicture}

\begin{axis}[%
width=0.951\figurewidth,
height=\figureheight,
at={(0\figurewidth,0\figureheight)},
scale only axis,
xmin=0.5,
xmax=5.5,
xtick={1,2,3,4,5},
xticklabels={{$\text{0.4 mm}^{\text{1}}$},{$\text{0.6 mm}^{\text{1,2}}$},{$\text{0.8 mm}^{\text{3}}$},{$\text{1.0 mm}^{\text{4}}$},{$\text{Rigid}^{\text{2,3,4}}$}},
xlabel style={font=\color{white!15!black}},
xlabel={Condition (-)},
ymin=0,
ymax=50,
ylabel style={font=\color{white!15!black}},
ylabel={Peak hammer velocity (rad/s)},
axis background/.style={fill=white},
axis x line*=bottom,
axis y line*=left
]
\addplot [color=mycolor1, mark=o, mark options={solid, mycolor1}, forget plot]
  table[row sep=crcr]{%
1	22.0414725713657\\
2	35.04993568133\\
3	34.230605969648\\
4	29.2089007859277\\
5	19.5311426677813\\
};
\addplot [color=mycolor2, mark=o, mark options={solid, mycolor2}, forget plot]
  table[row sep=crcr]{%
1	14.3312611099918\\
2	30.0184799430476\\
3	29.6993804224176\\
4	24.8647119203578\\
5	17.7991517483407\\
};
\addplot [color=mycolor3, mark=o, mark options={solid, mycolor3}, forget plot]
  table[row sep=crcr]{%
1	18.2185745019796\\
2	27.0594919031051\\
3	24.6542741466824\\
};
\addplot [color=mycolor4, mark=o, mark options={solid, mycolor4}, forget plot]
  table[row sep=crcr]{%
1	21.6220687524615\\
2	31.6023180320796\\
3	23.7698253578623\\
4	17.7811782451923\\
5	16.9509862539952\\
};
\addplot [color=mycolor5, mark=o, mark options={solid, mycolor5}, forget plot]
  table[row sep=crcr]{%
1	19.5935318418776\\
2	25.1206789387975\\
3	16.9300798087093\\
4	18.2527112564045\\
5	11.9575989982922\\
};
\addplot [color=mycolor6, mark=o, mark options={solid, mycolor6}, forget plot]
  table[row sep=crcr]{%
3	38.6246710104776\\
4	21.5813898534104\\
5	20.4077081567697\\
};
\addplot [color=mycolor7, mark=o, mark options={solid, mycolor7}, forget plot]
  table[row sep=crcr]{%
1	19.9046063512439\\
2	28.2752473088033\\
3	22.1979518248802\\
4	26.2802828183503\\
5	15.2667697739848\\
};
\addplot [color=mycolor1, mark=o, mark options={solid, mycolor1}, forget plot]
  table[row sep=crcr]{%
1	24.3553966019421\\
2	26.3686307276078\\
3	29.8895638930465\\
4	23.9326390465205\\
5	20.6093760992424\\
};
\addplot [color=mycolor2, mark=o, mark options={solid, mycolor2}, forget plot]
  table[row sep=crcr]{%
1	21.7502524958486\\
2	33.5599541390798\\
3	33.0518923112179\\
4	36.2936537383688\\
5	21.1651140213962\\
};
\addplot [color=mycolor3, mark=o, mark options={solid, mycolor3}, forget plot]
  table[row sep=crcr]{%
1	10.6944617676813\\
2	24.0255228177708\\
3	20.1199632565173\\
4	18.6868707465127\\
5	12.4587080608687\\
};
\addplot [color=mycolor4, mark=o, mark options={solid, mycolor4}, forget plot]
  table[row sep=crcr]{%
1	33.6044503719393\\
2	45.0911363778519\\
3	46.0625592873019\\
4	35.666098701703\\
5	18.7345447031225\\
};
\addplot [color=mycolor5, mark=o, mark options={solid, mycolor5}, forget plot]
  table[row sep=crcr]{%
1	29.2552007622831\\
2	34.3894727166386\\
3	20.8077603237032\\
4	19.6334487156102\\
5	17.8135422734748\\
};
\addplot [color=mycolor6, mark=o, mark options={solid, mycolor6}, forget plot]
  table[row sep=crcr]{%
1	21.3209730558166\\
2	16.1993124942071\\
3	29.2288450470094\\
4	22.413007706356\\
5	16.7566609243786\\
};
\addplot [color=black, line width=2.0pt, mark=o, mark options={solid, black}, forget plot]
  table[row sep=crcr]{%
1	21.471520904139\\
2	29.1468636259255\\
3	29.2288450470094\\
4	23.1728233764382\\
5	17.8063470109077\\
};
\end{axis}
\end{tikzpicture}%
		\caption{Peak hammer velocity by condition. The thin lines connect the median of the performance trials of each subject. The thick line connects the medians of the medians of all subjects. Superscript numbers identify pairs of conditions with statistically significant difference at the 0.5\%-level.}
		\label{fig:stats-hammer-velocity-friedman}
	\end{figure}
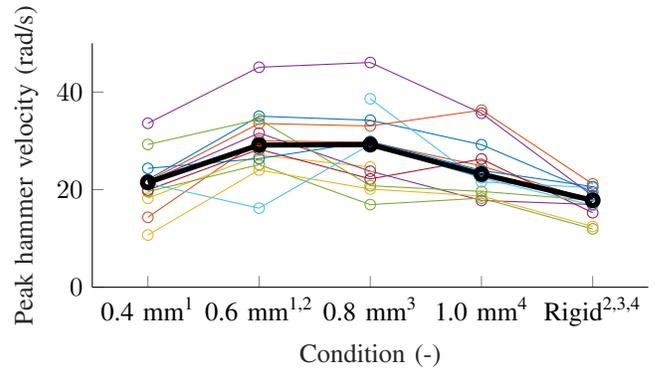

	For the following results, we only considered the trials of the performance phase. We calculated the median per participant for each condition and used Friedman's test to check for column effects after adjusting for possible row effects ($\alpha = 5\%$, row effects: subject-dependent variations, column effects: condition-dependent variations). We then used Wilcoxon's signed-rank test for pairwise testing of the conditions ($\alpha = \frac{2}{k\;(k+1)} \,\cdot\, 5\%$, with $k$ the number of conditions). With two participants, not all conditions could be performed for organizational reasons (participant~3 missing 1.0~mm condition and rigid condition, participant~6 missing 0.4~mm condition and 0.6~mm condition). Therefore, the results of these participants were excluded from Friedman's test ($n = 11$) and the concerned conditions were excluded from the pairwise testing ($n = 12$ for the tests including one of the conditions mentioned above, $n = 11$ for the tests including two of the conditions mentioned above).
	Fig.~\ref{fig:stats-hammer-velocity-friedman} shows the achieved peak output velocity results, ranging from 10.69~rad/s to 46.06~rad/s median per participant per condition. The column effects after adjusting for possible row effects are different at the 5\%-level (Friedman’s test: $\chi^2 = 25.60$, $n = 11$). The peak output velocity is significantly higher at the 0.5\%-level with the 0.6~mm, 0.8~mm and 1.0~mm leaf-springs compared to the rigid extension. The increase of median peak output velocity is more than 63\% from the rigid extension to the 0.6 mm and 0.8 mm leaf-springs (rank-biserial correlation for 0.6 mm vs rigid: 0.97, rank-biserial correlation for 0.8 mm vs rigid: 1.00, rank-biserial correlation for 1.0~mm vs rigid: 1.00).%
	
\begin{figure}[!t]
		\centering
		\setlength{\figurewidth}{.9\columnwidth}
		\setlength{\figureheight}{\plotheightA}
%
%
\definecolor{mycolor1}{rgb}{0.00000,0.44700,0.74100}%
\definecolor{mycolor2}{rgb}{0.85000,0.32500,0.09800}%
\definecolor{mycolor3}{rgb}{0.92900,0.69400,0.12500}%
\definecolor{mycolor4}{rgb}{0.49400,0.18400,0.55600}%
\definecolor{mycolor5}{rgb}{0.46600,0.67400,0.18800}%
\definecolor{mycolor6}{rgb}{0.30100,0.74500,0.93300}%
\definecolor{mycolor7}{rgb}{0.63500,0.07800,0.18400}%
\begin{tikzpicture}

\begin{axis}[%
width=0.951\figurewidth,
height=\figureheight,
at={(0\figurewidth,0\figureheight)},
scale only axis,
xmin=0.5,
xmax=5.5,
xtick={1,2,3,4,5},
xticklabels={{$\text{0.4 mm}^{\text{1,3}}$},{$\text{0.6 mm}^{\text{1,2,4}}$},{$\text{0.8 mm}^{\text{5}}$},{$\text{1.0 mm}^{\text{2,6}}$},{$\text{Rigid}^{\text{3,4,5,6}}$}},
xlabel style={font=\color{white!15!black}},
xlabel={Condition (-)},
ymin=0,
ymax=4,
ylabel style={font=\color{white!15!black}},
ylabel={Gain (-)},
axis background/.style={fill=white},
axis x line*=bottom,
axis y line*=left
]
\addplot [color=mycolor1, mark=o, mark options={solid, mycolor1}, forget plot]
  table[row sep=crcr]{%
1	1.46955949853994\\
2	2.13818977988945\\
3	2.48045205190791\\
4	1.57536452945621\\
5	1.03610094333053\\
};
\addplot [color=mycolor2, mark=o, mark options={solid, mycolor2}, forget plot]
  table[row sep=crcr]{%
1	1.88620288144218\\
2	1.95621910894669\\
3	2.61903208211674\\
4	1.54446772709709\\
5	1.00301310803389\\
};
\addplot [color=mycolor3, mark=o, mark options={solid, mycolor3}, forget plot]
  table[row sep=crcr]{%
1	1.66060594216684\\
2	1.7520133859636\\
3	2.57325021403115\\
};
\addplot [color=mycolor4, mark=o, mark options={solid, mycolor4}, forget plot]
  table[row sep=crcr]{%
1	1.64277667826252\\
2	2.25789683531717\\
3	1.65956535466813\\
4	1.23975135924348\\
5	1.00110163908249\\
};
\addplot [color=mycolor5, mark=o, mark options={solid, mycolor5}, forget plot]
  table[row sep=crcr]{%
1	1.75226414959474\\
2	2.22465117650274\\
3	1.37988603929045\\
4	1.26952124687889\\
5	0.987957594066498\\
};
\addplot [color=mycolor6, mark=o, mark options={solid, mycolor6}, forget plot]
  table[row sep=crcr]{%
3	2.17677166448531\\
4	1.19383832619013\\
5	0.994506434324647\\
};
\addplot [color=mycolor7, mark=o, mark options={solid, mycolor7}, forget plot]
  table[row sep=crcr]{%
1	1.79681255194637\\
2	3.7566443614626\\
3	2.03321829563721\\
4	3.13676769270845\\
5	1.01135882204275\\
};
\addplot [color=mycolor1, mark=o, mark options={solid, mycolor1}, forget plot]
  table[row sep=crcr]{%
1	1.81565199939071\\
2	2.47358535787819\\
3	2.28701731178827\\
4	2.31117997165196\\
5	1.00917494821298\\
};
\addplot [color=mycolor2, mark=o, mark options={solid, mycolor2}, forget plot]
  table[row sep=crcr]{%
1	1.46051831818769\\
2	1.5538992783469\\
3	1.51292286170631\\
4	1.71015579846997\\
5	1.01439976268443\\
};
\addplot [color=mycolor3, mark=o, mark options={solid, mycolor3}, forget plot]
  table[row sep=crcr]{%
1	1.04317634552675\\
2	1.75328955990795\\
3	1.66696676961409\\
4	1.62210663254562\\
5	1.0152148953404\\
};
\addplot [color=mycolor4, mark=o, mark options={solid, mycolor4}, forget plot]
  table[row sep=crcr]{%
1	1.69860872918071\\
2	2.11997214385941\\
3	2.39461915403082\\
4	1.66802713702214\\
5	1.00248367354159\\
};
\addplot [color=mycolor5, mark=o, mark options={solid, mycolor5}, forget plot]
  table[row sep=crcr]{%
1	1.62228397111374\\
2	1.76677732622733\\
3	1.20304619290754\\
4	1.07077271783609\\
5	1.00488830456792\\
};
\addplot [color=mycolor6, mark=o, mark options={solid, mycolor6}, forget plot]
  table[row sep=crcr]{%
1	1.84117441698721\\
2	1.71641991696397\\
3	2.05991009839103\\
4	1.66533572072285\\
5	1.0094038421177\\
};
\addplot [color=black, line width=2.0pt, mark=o, mark options={solid, black}, forget plot]
  table[row sep=crcr]{%
1	1.67960733567378\\
2	2.03809562640305\\
3	2.05991009839103\\
4	1.59873558100091\\
5	1.00703162639045\\
};
\end{axis}
\end{tikzpicture}%
		\caption{Gain by condition. The thin lines connect the median of the performance trials of each subject. The thick line connects the medians of the medians of all subjects. Superscript numbers identify pairs of conditions with statistically significant difference at the 0.5\%-level.}
		\label{fig:stats-gain-friedman}
	\end{figure}
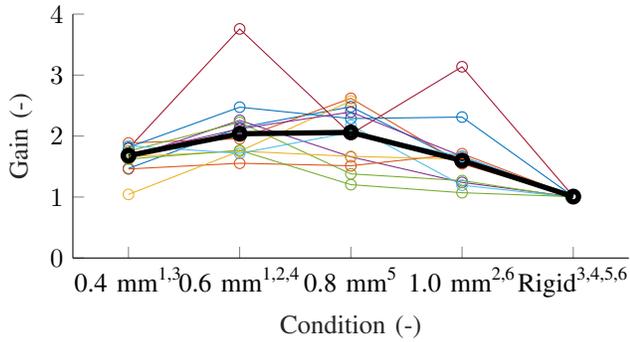

	Fig.~\ref{fig:stats-gain-friedman} shows the achieved gain results, ranging from 1.04 to 3.76 median per participant per condition (the gain for the rigid extension is always 1). After adjusting for possible row effects, the column effects are different at the 5\%-level (Friedman’s test: $\chi^2 = 30.84$, $n = 11$). The gain is significantly higher at the 0.5\%-level with the flexible extension compared to the rigid extension. The increase in median gain is more than 100\% from the rigid extension to the 0.6 mm leaf-spring (rank-biserial correlation of all flexible conditions vs the rigid condition: 1.00).%
	
\begin{figure}[!t]
		\centering
		\setlength{\figurewidth}{.9\columnwidth}
		\setlength{\figureheight}{\plotheightA}
%
%
\definecolor{mycolor1}{rgb}{0.00000,0.44700,0.74100}%
\definecolor{mycolor2}{rgb}{0.85000,0.32500,0.09800}%
\definecolor{mycolor3}{rgb}{0.92900,0.69400,0.12500}%
\definecolor{mycolor4}{rgb}{0.49400,0.18400,0.55600}%
\definecolor{mycolor5}{rgb}{0.46600,0.67400,0.18800}%
\definecolor{mycolor6}{rgb}{0.30100,0.74500,0.93300}%
\definecolor{mycolor7}{rgb}{0.63500,0.07800,0.18400}%
\begin{tikzpicture}

\begin{axis}[%
width=0.951\figurewidth,
height=\figureheight,
at={(0\figurewidth,0\figureheight)},
scale only axis,
xmin=0.5,
xmax=4.5,
xtick={1,2,3,4},
xticklabels={{$\text{0.4 mm}^{\text{1}}$},{$\text{0.6 mm}^{\text{1,2}}$},{0.8 mm},{$\text{1.0 mm}^{\text{2}}$}},
xlabel style={font=\color{white!15!black}},
xlabel={Condition (-)},
ymin=0,
ymax=110,
ylabel style={font=\color{white!15!black}},
ylabel={Relative gain (\%)},
axis background/.style={fill=white},
axis x line*=bottom,
axis y line*=left
]
\addplot [color=mycolor1, mark=o, mark options={solid, mycolor1}, forget plot]
  table[row sep=crcr]{%
1	39.5186926530225\\
2	59.0943896471358\\
3	70.1683542556132\\
4	41.7639786264492\\
};
\addplot [color=mycolor2, mark=o, mark options={solid, mycolor2}, forget plot]
  table[row sep=crcr]{%
1	50.72286765321\\
2	54.0651607946818\\
3	74.0885802664193\\
4	40.9448835095893\\
};
\addplot [color=mycolor3, mark=o, mark options={solid, mycolor3}, forget plot]
  table[row sep=crcr]{%
1	44.6562224336444\\
2	48.421408927735\\
3	72.7934782966627\\
};
\addplot [color=mycolor4, mark=o, mark options={solid, mycolor4}, forget plot]
  table[row sep=crcr]{%
1	44.1767663781632\\
2	62.4028028869197\\
3	46.946672332217\\
4	32.8666466087241\\
};
\addplot [color=mycolor5, mark=o, mark options={solid, mycolor5}, forget plot]
  table[row sep=crcr]{%
1	47.1210512017675\\
2	61.4839733543245\\
3	39.0349542789317\\
4	33.655866454461\\
};
\addplot [color=mycolor6, mark=o, mark options={solid, mycolor6}, forget plot]
  table[row sep=crcr]{%
3	61.5776810399144\\
4	31.6494610651488\\
};
\addplot [color=mycolor7, mark=o, mark options={solid, mycolor7}, forget plot]
  table[row sep=crcr]{%
1	48.3190255760387\\
2	103.824556524381\\
3	57.5167665658077\\
4	83.157831996911\\
};
\addplot [color=mycolor1, mark=o, mark options={solid, mycolor1}, forget plot]
  table[row sep=crcr]{%
1	48.8256470051439\\
2	68.3639115380401\\
3	64.6963688731032\\
4	61.2709434122327\\
};
\addplot [color=mycolor2, mark=o, mark options={solid, mycolor2}, forget plot]
  table[row sep=crcr]{%
1	39.275561546105\\
2	42.9460145636757\\
3	42.7983710630363\\
4	45.3373862872564\\
};
\addplot [color=mycolor3, mark=o, mark options={solid, mycolor3}, forget plot]
  table[row sep=crcr]{%
1	28.0526004035451\\
2	48.4566792863507\\
3	47.1560475166803\\
4	43.0031433771341\\
};
\addplot [color=mycolor4, mark=o, mark options={solid, mycolor4}, forget plot]
  table[row sep=crcr]{%
1	45.6781752443005\\
2	58.5908982862966\\
3	67.7402673347664\\
4	44.2205269931884\\
};
\addplot [color=mycolor5, mark=o, mark options={solid, mycolor5}, forget plot]
  table[row sep=crcr]{%
1	43.6256862781432\\
2	48.8294484978789\\
3	34.0324141258333\\
4	28.3869085950081\\
};
\addplot [color=mycolor6, mark=o, mark options={solid, mycolor6}, forget plot]
  table[row sep=crcr]{%
1	49.5119836779771\\
2	47.4376916049134\\
3	58.2718385575887\\
4	44.1491757276897\\
};
\addplot [color=black, line width=2.0pt, mark=o, mark options={solid, black}, forget plot]
  table[row sep=crcr]{%
1	45.1671988389724\\
2	56.3280295404892\\
3	58.2718385575887\\
4	42.3835610017917\\
};
\end{axis}
\end{tikzpicture}%
		\caption{Relative gain by condition. The thin lines connect the median of the performance trials of each subject. The thick line connects the medians of the medians of all subjects. Superscript numbers identify pairs of conditions with statistically significant difference at the 0.83\%-level.}
		\label{fig:stats-relgain-friedman}
	\end{figure}
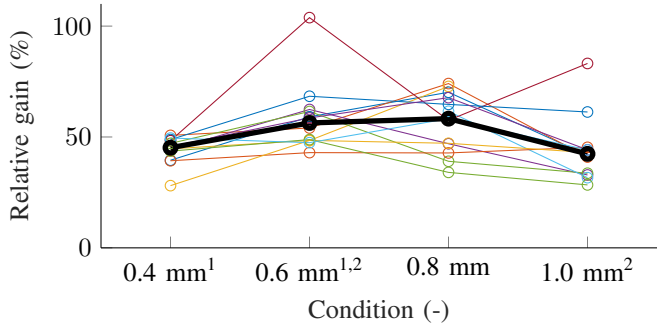

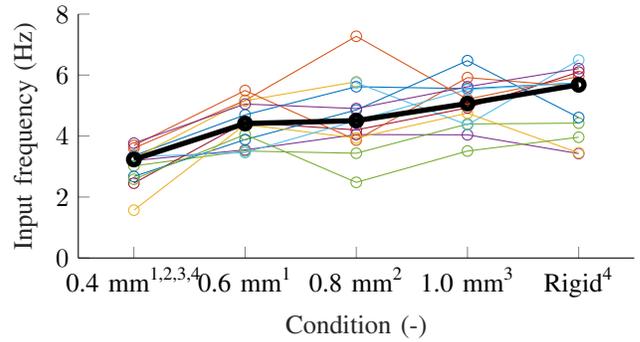
\begin{figure}[!t]
		\centering
		\setlength{\figurewidth}{.9\columnwidth}
		\setlength{\figureheight}{\plotheightA}
%
%
\definecolor{mycolor1}{rgb}{0.00000,0.44700,0.74100}%
\definecolor{mycolor2}{rgb}{0.85000,0.32500,0.09800}%
\definecolor{mycolor3}{rgb}{0.92900,0.69400,0.12500}%
\definecolor{mycolor4}{rgb}{0.49400,0.18400,0.55600}%
\definecolor{mycolor5}{rgb}{0.46600,0.67400,0.18800}%
\definecolor{mycolor6}{rgb}{0.30100,0.74500,0.93300}%
\definecolor{mycolor7}{rgb}{0.63500,0.07800,0.18400}%
\begin{tikzpicture}

\begin{axis}[%
width=0.951\figurewidth,
height=\figureheight,
at={(0\figurewidth,0\figureheight)},
scale only axis,
xmin=0.5,
xmax=5.5,
xtick={1,2,3,4,5},
xticklabels={{$\text{0.4 mm}^{\text{1,2,3,4}}$},{$\text{0.6 mm}^{\text{1}}$},{$\text{0.8 mm}^{\text{2}}$},{$\text{1.0 mm}^{\text{3}}$},{$\text{Rigid}^{\text{4}}$}},
xlabel style={font=\color{white!15!black}},
xlabel={Condition (-)},
ymin=0,
ymax=8,
ylabel style={font=\color{white!15!black}},
ylabel={Input frequency (Hz)},
axis background/.style={fill=white},
axis x line*=bottom,
axis y line*=left
]
\addplot [color=mycolor1, mark=o, mark options={solid, mycolor1}, forget plot]
  table[row sep=crcr]{%
1	3.35570469798658\\
2	4.69483568075115\\
3	5.61815484644765\\
4	5.55555555555556\\
5	5.74731626754762\\
};
\addplot [color=mycolor2, mark=o, mark options={solid, mycolor2}, forget plot]
  table[row sep=crcr]{%
1	3.60370889823905\\
2	5.16826923076932\\
3	7.27282344229368\\
4	5.20847462224996\\
5	5.95238095238109\\
};
\addplot [color=mycolor3, mark=o, mark options={solid, mycolor3}, forget plot]
  table[row sep=crcr]{%
1	3.26797385620914\\
2	5.15957446808508\\
3	5.78034682080937\\
};
\addplot [color=mycolor4, mark=o, mark options={solid, mycolor4}, forget plot]
  table[row sep=crcr]{%
1	3.1949207904622\\
2	3.54609929078019\\
3	4.0485829959514\\
4	4.04055245371729\\
5	3.42465753424656\\
};
\addplot [color=mycolor5, mark=o, mark options={solid, mycolor5}, forget plot]
  table[row sep=crcr]{%
1	3.02114803625375\\
2	3.50877192982456\\
3	3.43642611683846\\
4	4.39562562794651\\
5	4.42477876106193\\
};
\addplot [color=mycolor6, mark=o, mark options={solid, mycolor6}, forget plot]
  table[row sep=crcr]{%
3	5.74712643678163\\
4	4.36681222707452\\
5	6.4935064935065\\
};
\addplot [color=mycolor7, mark=o, mark options={solid, mycolor7}, forget plot]
  table[row sep=crcr]{%
1	2.45416877454169\\
2	4.44479563817392\\
3	4.20168067226879\\
4	4.92610837438421\\
5	6.0975609756097\\
};
\addplot [color=mycolor1, mark=o, mark options={solid, mycolor1}, forget plot]
  table[row sep=crcr]{%
1	2.67379679144386\\
2	3.88350978795282\\
3	4.85436893203884\\
4	6.47255969836628\\
5	4.6082949308756\\
};
\addplot [color=mycolor2, mark=o, mark options={solid, mycolor2}, forget plot]
  table[row sep=crcr]{%
1	3.69003690036904\\
2	5.49450549450543\\
3	3.86848642662583\\
4	5.91715976331378\\
5	5.61797752808997\\
};
\addplot [color=mycolor3, mark=o, mark options={solid, mycolor3}, forget plot]
  table[row sep=crcr]{%
1	1.56985871271586\\
2	4.3768915450332\\
3	3.93700787401578\\
4	4.75083506364536\\
5	3.4489320203606\\
};
\addplot [color=mycolor4, mark=o, mark options={solid, mycolor4}, forget plot]
  table[row sep=crcr]{%
1	3.76649170095047\\
2	5.05050505050522\\
3	4.90196078431371\\
4	5.61797752808986\\
5	6.21118012422355\\
};
\addplot [color=mycolor5, mark=o, mark options={solid, mycolor5}, forget plot]
  table[row sep=crcr]{%
1	2.58397932816542\\
2	4.04884845504458\\
3	2.48138957816378\\
4	3.50881512853347\\
5	3.96165432408689\\
};
\addplot [color=mycolor6, mark=o, mark options={solid, mycolor6}, forget plot]
  table[row sep=crcr]{%
1	3.32225913621261\\
2	3.46020761245665\\
3	4.50459590528169\\
4	5.52486187845312\\
5	5.74712643678139\\
};
\addplot [color=black, line width=2.0pt, mark=o, mark options={solid, black}, forget plot]
  table[row sep=crcr]{%
1	3.23144732333567\\
2	4.41084359160356\\
3	4.50459590528169\\
4	5.06729149831708\\
5	5.68255198243568\\
};
\end{axis}
\end{tikzpicture}%
		\caption{Input frequency by condition. The thin lines connect the median of the performance trials of each subject. The thick line connects the medians of the medians of all subjects. Superscript numbers identify pairs of conditions with statistically significant difference at the 0.5\%-level.}
		\label{fig:stats-efreq-friedman}
	\end{figure}

	Fig.~\ref{fig:stats-relgain-friedman} shows the achieved relative gain results, ranging from 28.05\% to 103.82\% median per participant per condition (the latter being an unusual subject as the others achieved at most 68.36\% for this condition). After adjusting for possible row effects, the column effects are different at the 5\%-level (Friedman’s test: $\chi^2 = 14.67$, $n = 11$). The relative gain is significantly higher at the 0.83\%-level for the 0.6~mm leaf-spring compared to the 0.4~mm and the 1.0~mm leaf-spring (rank-biserial correlation for 0.6 mm vs 0.4 mm: 0.97, rank-biserial correlation for 0.6 mm vs 1.0 mm: 0.97). There is no significant difference between the 0.6~mm and the 0.8~mm leaf-spring.%
	Fig.~\ref{fig:stats-efreq-friedman} shows the estimated input frequency results, ranging from 1.57~Hz to 7.27~Hz median per participant per condition. After adjusting for possible row effects, the column effects are different at the 5\%-level (Friedman’s test: $\chi^2 = 24.87$, $n = 11$). The estimated input frequency is significantly lower at the 0.83\%-level for the 0.4~mm leaf-spring compared to all other conditions (rank-biserial correlation for 0.4 mm vs 0.6 mm: -1.00, rank-biserial correlation for 0.4 mm vs 0.8 mm: -0.97, rank-biserial correlation for 0.4 mm vs 1.0 mm: -1.00, rank-biserial correlation for 0.4 mm vs rigid: -1.00).%
	
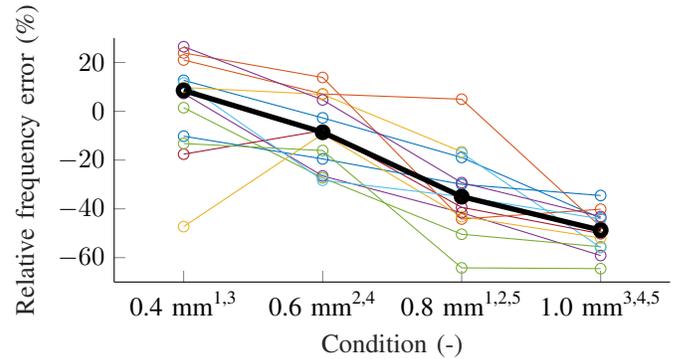
\begin{figure}[!t]
		\centering
		\setlength{\figurewidth}{.9\columnwidth}
		\setlength{\figureheight}{\plotheightA}
%
%
\definecolor{mycolor1}{rgb}{0.00000,0.44700,0.74100}%
\definecolor{mycolor2}{rgb}{0.85000,0.32500,0.09800}%
\definecolor{mycolor3}{rgb}{0.92900,0.69400,0.12500}%
\definecolor{mycolor4}{rgb}{0.49400,0.18400,0.55600}%
\definecolor{mycolor5}{rgb}{0.46600,0.67400,0.18800}%
\definecolor{mycolor6}{rgb}{0.30100,0.74500,0.93300}%
\definecolor{mycolor7}{rgb}{0.63500,0.07800,0.18400}%
\begin{tikzpicture}

\begin{axis}[%
width=0.951\figurewidth,
height=\figureheight,
at={(0\figurewidth,0\figureheight)},
scale only axis,
xmin=0.5,
xmax=4.5,
xtick={1,2,3,4},
xticklabels={{$\text{0.4 mm}^{\text{1,3}}$},{$\text{0.6 mm}^{\text{2,4}}$},{$\text{0.8 mm}^{\text{1,2,5}}$},{$\text{1.0 mm}^{\text{3,4,5}}$}},
xlabel style={font=\color{white!15!black}},
xlabel={Condition (-)},
ymin=-70,
ymax=30,
ylabel style={font=\color{white!15!black}},
ylabel={Relative frequency error (\%)},
axis background/.style={fill=white},
axis x line*=bottom,
axis y line*=left
]
\addplot [color=mycolor1, mark=o, mark options={solid, mycolor1}, forget plot]
  table[row sep=crcr]{%
1	12.6979537840369\\
2	-2.70467064070532\\
3	-18.9210509998412\\
4	-43.8023632718987\\
};
\addplot [color=mycolor2, mark=o, mark options={solid, mycolor2}, forget plot]
  table[row sep=crcr]{%
1	21.0269244217305\\
2	7.10672134637013\\
3	4.95846004275544\\
4	-47.3132863488272\\
};
\addplot [color=mycolor3, mark=o, mark options={solid, mycolor3}, forget plot]
  table[row sep=crcr]{%
1	9.75160205112057\\
2	6.9265319091695\\
3	-16.5803617207225\\
};
\addplot [color=mycolor4, mark=o, mark options={solid, mycolor4}, forget plot]
  table[row sep=crcr]{%
1	7.29818860497607\\
2	-26.5109746328719\\
3	-41.5724800716005\\
4	-59.1274901845284\\
};
\addplot [color=mycolor5, mark=o, mark options={solid, mycolor5}, forget plot]
  table[row sep=crcr]{%
1	1.46220612580871\\
2	-27.2845433209479\\
3	-50.4068817102591\\
4	-55.5357209982274\\
};
\addplot [color=mycolor6, mark=o, mark options={solid, mycolor6}, forget plot]
  table[row sep=crcr]{%
3	-17.0597849292255\\
4	-55.8271851045463\\
};
\addplot [color=mycolor7, mark=o, mark options={solid, mycolor7}, forget plot]
  table[row sep=crcr]{%
1	-17.5792198588053\\
2	-7.88647676766042\\
3	-39.3630360406962\\
4	-50.1695831967578\\
};
\addplot [color=mycolor1, mark=o, mark options={solid, mycolor1}, forget plot]
  table[row sep=crcr]{%
1	-10.2032346854462\\
2	-19.5185114916609\\
3	-29.9437018334237\\
4	-34.5263394466474\\
};
\addplot [color=mycolor2, mark=o, mark options={solid, mycolor2}, forget plot]
  table[row sep=crcr]{%
1	23.9261632016359\\
2	13.8676107336792\\
3	-44.1715612572472\\
4	-40.1445289286478\\
};
\addplot [color=mycolor3, mark=o, mark options={solid, mycolor3}, forget plot]
  table[row sep=crcr]{%
1	-47.2778803333702\\
2	-9.29371475344525\\
3	-43.1826873137211\\
4	-51.9425734488638\\
};
\addplot [color=mycolor4, mark=o, mark options={solid, mycolor4}, forget plot]
  table[row sep=crcr]{%
1	26.4938204772191\\
2	4.66618764409374\\
3	-29.2568753808104\\
4	-43.1709291513585\\
};
\addplot [color=mycolor5, mark=o, mark options={solid, mycolor5}, forget plot]
  table[row sep=crcr]{%
1	-13.2196634944611\\
2	-16.0920486366401\\
3	-64.189584560013\\
4	-64.5063187709098\\
};
\addplot [color=mycolor6, mark=o, mark options={solid, mycolor6}, forget plot]
  table[row sep=crcr]{%
1	11.5747183642621\\
2	-28.2909856279265\\
3	-34.9914853447674\\
4	-44.1128474527162\\
};
\addplot [color=black, line width=2.0pt, mark=o, mark options={solid, black}, forget plot]
  table[row sep=crcr]{%
1	8.52489532804832\\
2	-8.59009576055283\\
3	-34.9914853447674\\
4	-48.7414347727925\\
};
\end{axis}
\end{tikzpicture}%
		\caption{Relative frequency error by condition. The thin lines connect the median of the performance trials of each subject. The thick line connects the medians of the medians of all subjects. Superscript numbers identify pairs of conditions with significant difference at the 0.83\%-level.}
		\label{fig:stats-relative-frequency-error-friedman}
	\end{figure}

	Fig.~\ref{fig:stats-relative-frequency-error-friedman} shows the relative frequency error results, ranging from -64.51\% to 26.49\% median per participant per condition. After adjusting for possible row effects, the column effects are different at the 5\%-level (Friedman’s test: $\chi^2 = 27.76$, $n = 11$). The relative frequency error is significantly better at the 0.83\%-level for the 0.4~mm and 0.6~mm leaf-spring compared to the 0.8~mm and 1.0~mm conditions (rank-biserial correlation for 0.4 mm vs 0.8 mm: 0.97, rank-biserial correlation for 0.4 mm vs 1.0 mm: 1.00, rank-biserial correlation for 0.6 mm vs 0.8 mm: 1.00, rank-biserial correlation for 0.6 mm vs 1.0 mm: 1.00).%
	
\begin{figure*}[!t]
		\centering
		\setlength{\figurewidth}{.9\textwidth}
		\setlength{\figureheight}{\plotheightA}
		\input{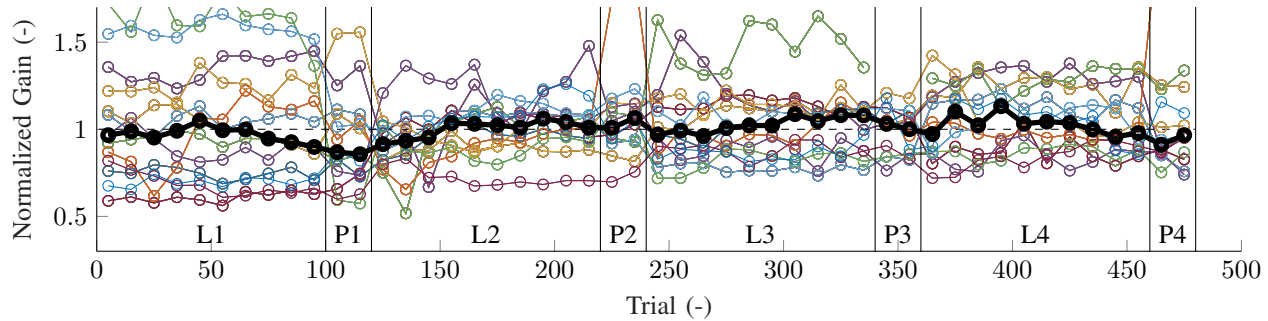}
		\caption{Normalized gain over cumulated trial count for all flexible condition trials. A value of 1 corresponds to the median over all trials of the same condition. Thin lines connect the median gains over ten trials per participant. The thick line connects the median of the medians over ten trials of all participants. Learning phases, labeled L1 to L4, consist of 100 trials and performance phases, labeled P1 to P4, consist of 20 trials. Vertical bars show the limits of the phases. After every learning phase (L1, L2, L3, L4) there was a short interruption to download the measurement data; after every performance phase (P1, P2, P3, P4) there was a short break to change the hammer extension.}
		\label{fig:gain-learning-normalized-friedman}
	\end{figure*}

	Fig.~\ref{fig:gain-learning-normalized-friedman} shows the gain over the cumulative trial count, independent of the condition, combining learning and performance trials. We normalized the gains so that a value of 1 corresponds to the median over all trials of the same condition, and we grouped per ten trials, calculating the medians per participant and the medians of the medians of all participants. No significant learning effect can be seen for most learning phases. In some cases, a small learning effect can be observed over the first 20 to 30 trials of the learning phase, and in others, the performance decreases after 60 to 70 trials, perhaps due to participants getting bored.
	
	\section{Discussion}
	\label{sec:discussion}
	The participants were able to use the elastic tool to their benefit, consistently reaching higher peak output velocities in the hammering motion in the flexible conditions than in the rigid condition, exploiting the spring gain. They achieved this performance after a short learning phase of less than 20 to 30 trials for new stiffness conditions, not showing significant improvements over the longer learning phase of 100 trials. Therefore, the participants seem to have an intuitive understanding of the dynamic system properties.
	
	The participants actively adapted to the system resonance by adopting higher excitation frequencies for stiffer springs, consistent with Hatsopoulos et al. \cite{hatsopoulos_resonance_1996}. However, the participants stayed below approximately 6~Hz excitation frequency (one outlier for the 0.8~mm condition), even for the 0.8~mm and 1.0~mm conditions that would require higher excitation frequencies for best performance (resonance frequency of 6.93~Hz and 9.89~Hz). We conclude that human dynamic limitations only allow input frequencies of up to approximately 6~Hz for the task used in our experiment. Overall, the participants achieve around half to two-thirds of the gain possible with a perfect velocity source (unexplained outliers achieved a higher gain), another effect of the human dynamic limitations.
	
	The participants achieved the best results with the 0.6~mm leaf-spring, obtaining a median gain of 2.04. Although some participants performed better with the 0.8~mm leaf-spring, others had more difficulties with the faster motion, resulting in a bigger spread of results between participants. Therefore, we observed no statistically significant improvement in the gain for the 0.8~mm leaf-spring compared to the other flexible conditions, and the 0.6~mm leaf-spring with a resonance frequency of 4.82~Hz seems to be the one the participants could best adjust to.
	
	Although we could show that humans can use elastic tools intuitively, we cannot yet explain the feedback mechanisms important for humans to be able to find and excite the system resonance frequency. We intend to do follow-up experiments in a teleoperation setup to separate the effects of force and visual feedback on human performance.
	
	\section{Conclusion}
	\label{sec:conclusion}
	This paper studied how humans interact with elastic tools. We performed an experiment with untrained participants hammering with an elastic hammer. The participants were able to intuitively exploit the elastic properties to their advantage, achieving significantly higher peak output velocities with an elastic hammer than with a rigid hammer. The participants quickly adapted to the resonance frequency for different stiffnesses. We conclude that humans should be able to use series elastic actuators also in teleoperator systems, if the teleoperator is sufficiently transparent. Our future research will study the feedback mechanisms used by humans for dynamic tasks and how to develop dynamic teleoperator systems that are intuitive to use. With this basis, a novel class of teleoperator systems can be developed that allow force amplification for dynamic tasks requiring high peak forces on the slave side.
	

\end{document}